\def\cm{cm$^{-1}$}
\def\efa{EuFe$_2$As$_{2}$}

\def\bfca{Ba(Fe$_{0.92}$Co$_{0.08})_2$\-As$_{2}$}
\def\bfna{Ba(Fe$_{0.95}$Ni$_{0.05})_2$\-As$_{2}$}

\def\tc{$T_{c}$}

\documentclass{elsart3p}

\usepackage{epsfig}

\usepackage{amssymb}

\begin{document}

\begin{frontmatter}



\title{Superfluid density of Ba(Fe$_{1-x}M_x$)$_2$As$_2$ from 
optical experiments}

\author[a]{D. Wu},
\author[a]{N. Bari\v{s}i\'{c}},
\author[a]{N. Drichko},
\author[a]{P. Kallina},
\author[a]{A. Faridian},
\author[a]{B. Gorshunov},
\author[a]{M. Dressel},
\author[b]{L. J. Li},
\author[b]{X. Lin}
\author[b]{G. H. Cao},
\author[b]{Z. A. Xu},
\address[a]{1. Physikalisches Institut, Universit\"at Stuttgart, Pfaffenwaldring 57, 70550 Stuttgart, Germany}
\address[b]{Department of Physics, Zhejiang University, Hangzhou 310027, People's Republic of China}

\begin{abstract}
The temperature dependence of the $ab$-plane optical reflectivity of \bfca\
and \bfna\ single crystals is measured in a
wide spectral range. Upon entering the superconducting regime, the
reflectivity in both compounds increases considerably at low
frequency, leading to a clear gap in the optical
conductivity  below 100~\cm. From the analysis of the complex
conductivity spectra we obtain the penetration depth
$\lambda(T)=(3500\pm 350)$~\AA\ for \bfca\ and $(3000\pm 300)$~\AA\ for \bfna. The
calculated superfluid density $\rho_s$ of both compounds nicely fits
Homes' scaling relation $\rho_s=(125\pm 25)\sigma_{dc}T_c$.
 \end{abstract}

\end{frontmatter}

$A$Fe$_2$As$_2$ ($A$ = Ba, Sr, Eu) are ThCr$_2$Si$_2$-type ternary
iron arsenides which exhibit a spin-density-wave (SDW) instability
around 200~K. The development of the SDW gap was in detail
investigated in \efa, for instance \cite{Wu09}. 
Additional charge carriers are provided by partial 
replace of Ba by K \cite{rotter} or substitution
of Fe by Co or Ni. Concomitantly the SDW transition shifts
to lower temperatures and finally superconductivity sets in \cite{xu}. 
Here we present a
comprehensive optical study on optimal electron-doped \bfca\
(\tc=25~K) and \bfna\ (\tc=20~K) single crystals. The
superconducting gap is clearly observed in reflectivity spectra for $T<T_c$.
The spectral weight analysis on optical conductivity provides
information on the  penetration depth and the superfluid density.

Single crystals of \bfca\ and \bfna\ were synthesized using
self-flux method \cite{xu}. The platelets with a typical size of
2~mm $\times$ 2~mm $\times$ 0.1~mm have naturally flat and shiny
surfaces.
The resistivity and susceptibility evidence sharp superconducting
transitions at 25~K and 20~K, indicating that samples are
doped uniformly\cite{Barisic08}. The temperature dependence of optical
reflectivity was measured in a wide frequency range from 20 to
37\,000~\cm\ using a coherent-source spectrometer in the THz
range, infrared Fourier transform spectrometers (Bruker IFS 66v/s
and IFS 113v) and a Woollam spectroscopic ellipsometer extending
up to the ultraviolet. The low-frequency extra\-polation was done
according to the dc conductivity measured on the same crystals by
standard four-probe method. The optical conductivity was
calculated from the reflectivity spectra using Kramers-Kronig
analysis.

\begin{figure}
\centering
\includegraphics[width=7.5cm]{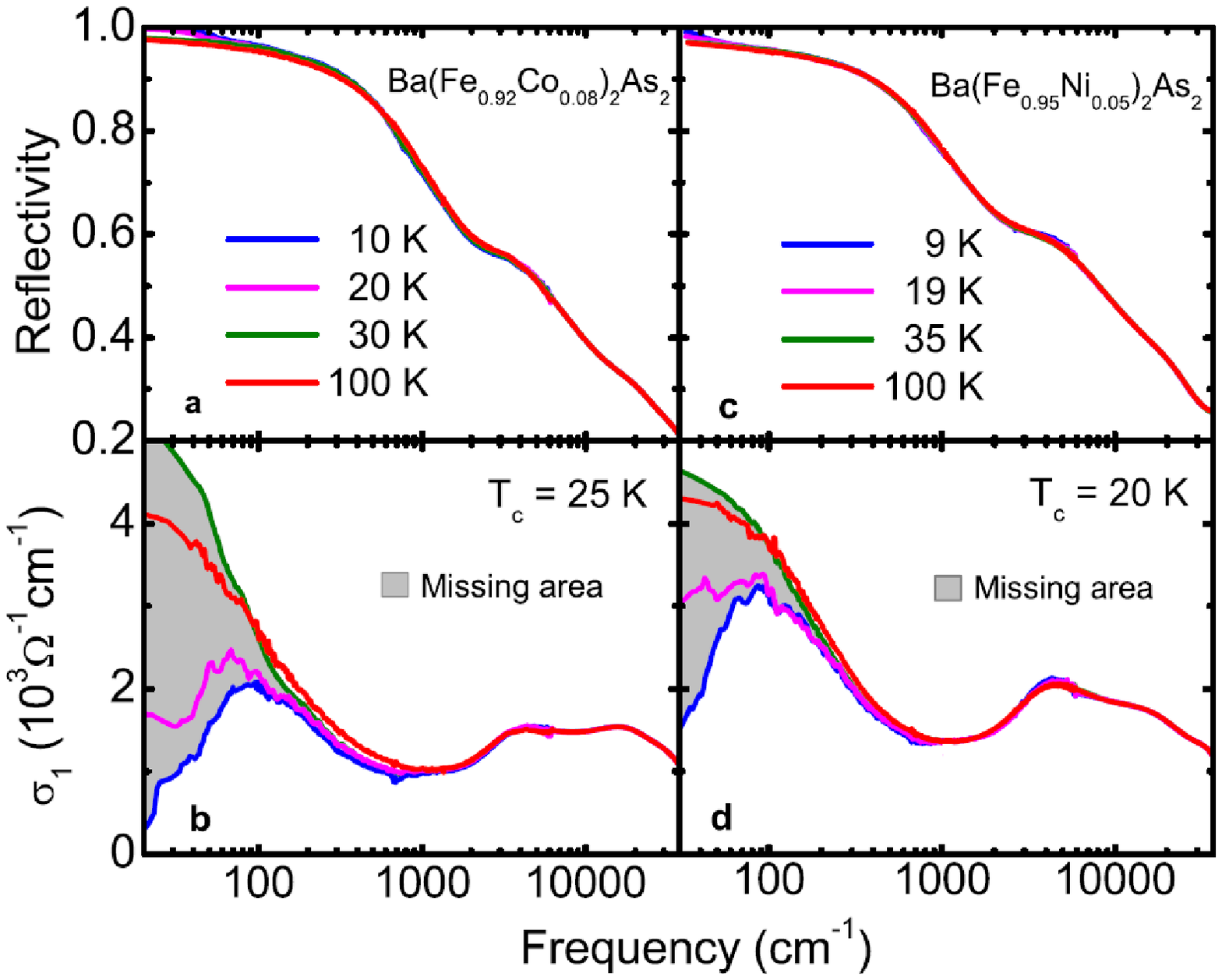}
\caption{\label{fig:optics} (a,c) $ab$-plane optical reflectivity
of \bfca\ and \bfna\ measured over a wide frequency range shown
for selected temperatures above and below the superconducting
transition. (b,d) Frequency dependent conductivity obtained from
the Kramers-Kronig analysis of the reflectivity. The missing
spectral weights for both materials, between the conductivity
curve just above the superconducting and the lowest measured
temperature (30~K and 10~K, 35~K and 9~K respectively) are
indicated by gray areas.}
\end{figure}

Fig.~\ref{fig:optics} shows the temperature dependent reflectivity
and conductivity of Ba(Fe$_{1-x}M_x$)$_2$As$_2$. The R($\omega$)
spectra show a good metallic behavior above \tc\ as the
reflectivity goes towards unity at low frequencies and increases
by cooling. Upon entering the superconducting regime, the
reflectivity starts to increase rapidly with a change in curvature
below 70~\cm\ and 60~\cm, respectively. This yields a gap-like
feature in the $\sigma(\omega)$, indicating a formation of
superconducting gap due to the pairing of electrons. According to
the Ferrell-Glover-Tinkham  sum rule, the removed spectral weight
in $\sigma^{(s)}_1(\omega)$ compared to the normal-state
$\sigma^{(n)}_1(\omega)$ is related to the formation of
superconducting condensate:
$A=\omega^2_{ps}/8={c^2}/{8\lambda^2}=\int\left[\sigma^{(n)}_1(\omega)-
\sigma^{(s)}_1(\omega)\right]{\rm d}\omega$, where the
${\omega}_{ps}$ is the plasma frequency of the Cooper pairs and
$\lambda$ is the penetration depth \cite{martin}. Thus, from the
missing area we directly estimate: $\lambda=(3500\pm 350)$~\AA\ for \bfca\
and $(3000\pm 300)$~\AA\ for \bfna. These values are in agreement with the
results from measurements of the microwave surface-impedance, muon
spin rotation and susceptibility. Another way to determine the
penetration depth is from the imaginary part of the optical
conductivity: $\lambda=c/\sqrt{4\pi\omega\sigma_2}$. Within the
low-frequency limit, $\lambda$ calculated from imaginary part of
the conductivity $\sigma_2(\omega)$ agrees well with the results
from the spectral weight analysis.

\begin{figure}
\centering
\includegraphics[width=7.5cm]{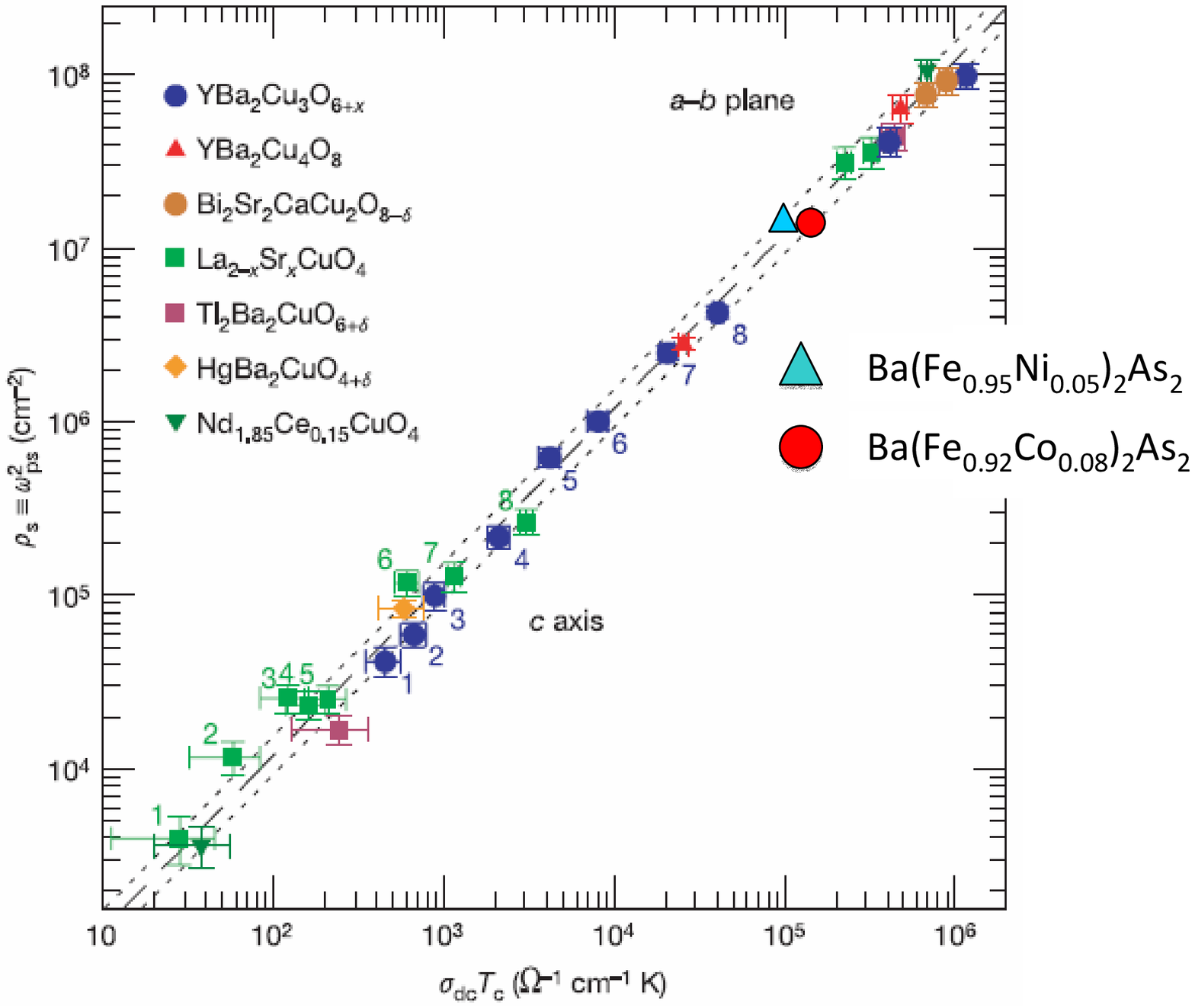}
\caption{\label{fig:Homes} Superfluid density $\rho_s$ versus
$\sigma_{dc}$\tc\ according to Homes {\it et al.} \cite{Homes}.
\bfca\ and \bfna\ fall right on the scaling relation
$\rho_s=(125\pm25)\sigma_{dc}T_c$ (lines) suggested for copper
oxides.}
\end{figure}

The missing area is also a measure of the superfluid density
$\rho_s=A/8$. Originally Uemura {\it et al.} \cite{Uemura91}
suggested a relation $\rho_s \propto T_c$ which works well in the
case of underdoped cuprates. More recently, Homes {\it et al.}
\cite{Homes} extended the scaling relation by an additional term
covering the optimal and overdoped cuprates as well:
$\rho_s=(125\pm25)\sigma_{dc}T_c$, where the $\sigma_{dc}$ is the
value of dc conductivity just above \tc. Interesting this relation
holds not only for the $ab$-plane but also for the measurements
performed along the $c$-axis. In Fig.~\ref{fig:Homes} we demonstrate
that Homes' relation also holds very well for
Ba(Fe$_{1-x}M_x$)$_2$As$_2$ family of materials. It is worth
noting that the iron-pnictides are distinctively different to
cuprates. While in iron pnictides the parent compounds are metallic, in cuprates they are insulating. Nevertheless they both fall right
on this scaling relation. Studies over a larger doping range
are needed to verify this scaling and draw further conclusions .


\end{document}